\documentclass[lettersize,journal]{IEEEtran}
\usepackage{amsmath,amsfonts}
\usepackage{algorithmic}
\usepackage{algorithm}
\usepackage{array}
\usepackage[caption=false,font=normalsize,labelfont=sf,textfont=sf]{subfig}
\usepackage{textcomp}
\usepackage{stfloats}
\usepackage{url}
\usepackage{verbatim}
\usepackage{graphicx}
\usepackage{cite}
\usepackage{color}
\usepackage[switch]{lineno}
\usepackage{fancyhdr}
\usepackage{bm}
\usepackage{flushend}

\begin{document}

\title{2D Slice-driven Physics-based 3D Motion Estimation Framework for Pancreatic Radiotherapy}

\author{Yuki Hara,~\IEEEmembership{Graduate Student Member,~IEEE}, Noriyuki Kadoya, Naoto Mitsume, Naoto Ienaga,~\IEEEmembership{Member,~IEEE},  Rei Umezawa, Keiichi Jingu, and Yoshihiro Kuroda,~\IEEEmembership{Member,~IEEE}
\thanks{Manuscript received April XX, 20XX; revised August XX, 20XX. This work was partly supported by AMED under Grant Number JP23YM0126803.} 
(\textit{Corresponding author: Yoshihiro Kuroda})%
\thanks{This work involved human subjects or animals in its research. The authors confirm that all human/animal subject research procedures and protocols are exempt from review board approval.}%
\thanks{Y. Hara is with the Degree Programs in Systems and Information Engineering, University of Tsukuba, Tsukuba, Japan (e-mail: hara.yuki.ss@alumni.tsukuba.ac.jp).}%
\thanks{N. Kadoya, R. Umezawa, and K. Jingu are with the Department of Radiation Oncology, Tohoku University Graduate School of Medicine, Sendai, Japan.}%
\thanks{N. Mitsume, N. Ienaga, and Y. Kuroda are 
with the Institute of Systems and Information Engineering, University of Tsukuba, Tsukuba, Japan (e-mail: kuroda@iit.tsukuba.ac.jp).}%
\thanks{Color versions of one or more of the figures in this article are available online at https://ieeexplore.ieee.org.}%
}



\maketitle
\thispagestyle{fancy}
\begin{abstract}
Pancreatic diseases are difficult to treat with high doses of radiation, as they often present both
periodic and aperiodic deformations. Nevertheless, we expect that these difficulties can be overcome, and treatment results may be improved with the practical use of a device that can capture 2D slices of organs during irradiation. However, since only a few 2D slices can be taken, the 3D motion needs to be estimated from partially observed information. In this study, we propose a physics-based framework for estimating the 3D motion of organs, regardless of periodicity, from motion information obtained by 2D slices in one or more directions and a regression model that estimates the accuracy of the proposed framework to select the optimal slice. 
Using information obtained by slice-to-slice registration and setting the surrounding organs as boundaries, the 
framework drives the physical models for  estimating 3D motion.
The R2 score of the proposed regression model 
was greater than 0.9, and the RMSE 
was 0.357 mm.

The mean errors were $\mathbf{5.11 \pm 1.09}$ mm using an axial slice and $\mathbf{2.13 \pm 0.598}$ mm using concurrent axial, sagittal, and coronal slices.
Our results suggest that the proposed framework is comparable to volume-to-volume registration, and 
is feasible.

\end{abstract}


\section{INTRODUCTION}
\label{sec:intro}
\IEEEPARstart{R}{adiotherapy} is recognized as the standard-of-care treatment for most cancers, and carries the advantage of being minimally invasive.
The results of radiotherapy treatment depend on the intensity of radiation that can be delivered to the target organ.
In areas with many organs in close proximity, such as in the abdomen, delivering high doses of radiation is difficult because the risk of irradiating non-target organs is high. This problem is particularly prominent in the case of the pancreas.
The pancreas is adjacent to many other organs such as the duodenum, stomach, kidney, and small bowel. It also undergoes both periodic motion due to respiration, and aperiodic motion due to contact. The five-year relative survival rate of patients with pancreatic cancer (10\%) was much lower than that of all cancers (67\%) in the United States between 2010 and 2016~\cite{cancer_statistics}; therefore, the treatment of pancreatic cancer remains challenging.
By contrast, it has been reported that increasing the radiation dose to the pancreas significantly increases the survival rate at 2 years~\cite{rudra}; therefore, precise irradiation is a viable approach.
\par A newly developed medical device, the MRI (Magnetic Resonance Imaging) linear accelerator (MR-Linac)~\cite{mr_linac}, is already in practical use and is expected to improve treatment results for pancreatic cancer. The MR-Linac can modify irradiation based on 2D slices captured during treatment.
However, because the MR-Linac can capture only a few 2D slices, out-of-plane motion of the pancreas remains undetected.
Therefore, an estimation method to predict 3D motion from a few 2D slices is required.
In addition, because pancreatic motion is complex, an estimation method that can handle aperiodic motion is required.
 \par To estimate 3D motion from 2D slices, a comprehensive survey of registration criteria, motion models, and optimization methods has been conducted for medical imaging tasks such as image fusion, motion correction, and volume reconstruction~\cite{ferrante}, and many methods have also been studied for MRI and radiotherapy.
 For deformable image registration (DIR), estimation methods based on DIR that apply the deformation obtained by specific slice-to-slice registration to all slices in the same direction~\cite{paganelli1} and use an optical flow algorithm~\cite{seregni} were explored.
In terms of statistical modeling methods, estimation methods used to go from 2D slices to 3D motion create displacement fields by Principal Component Analysis (PCA)~\cite{stemkens1,harris1,harris2,garau}.
These methods have been comprehensively compared~\cite{paganelli2}.
In terms of machine-learning methods, 3D motion estimation and future prediction using the sequence-to-sequence mechanism were reviewed~\cite{romaguera1,romaguera2, lombardo}.
A method that combines a statistical model and machine-learning methods uses machine-learning to optimize the PCA weights to generate deformation field maps~\cite{pham1}.
However, these methods were designed to express the periodic motion of organs caused by respiration.
In addition, these methods require a large amount of 4D-MRI or equivalent 4D data, and statistical models are known to have limitations with regard  to movements caused by irregular respiration.

\par An alternative approach to statistical modeling and machine-learning is the use of physical simulation.
Physical models can express 3D motion while considering the physical properties and boundary constraints of an object. Physical models have been used to estimate the 3D motion of pig livers~\cite{trivisonne} and the temporal motion of female pelvic organs~\cite{courtecuisse}, using the motion of each organ's contour obtained from 2D slices. However, these methods must accurately handle the contour of each organ to set proper boundary conditions. To the best of our knowledge, no studies thus far have investigated the various motions of the pancreas due to its contact with surrounding organs; therefore, a 3D estimation method of the pancreas has not yet been established.
\par The objective of this study was to model pancreatic motion, regardless of the presence or absence of both 
periodic motion with respiration and/or aperiodic motion with contact between organs.
We proposed a 2D slice-driven 3D motion framework using physical simulation to interpolate the out-of-plane motion that was not obtained from captured images.
In addition, because we hypothesized  that using the organs around the pancreas as boundaries would be a valid way to handle the aperiodic motion more accurately, we discretized not only the pancreas but the surrounding organs as well.
\par Since 3D motion is estimated using 2D information, the estimation accuracy is considered to be greatly affected by the driving slice's position and the combination of organs that it modeled.
However, it would be time-consuming to search for the optimal condition in a brute-force manner.
We therefore proposed a regression model to select the optimal slice position and combination of surrounding organs in the framework. Its input is the region of the organs in each slice, and its output is the accuracy of the proposed framework. 
\begin{figure}[tb]
  \centering
  \includegraphics[width=\linewidth]{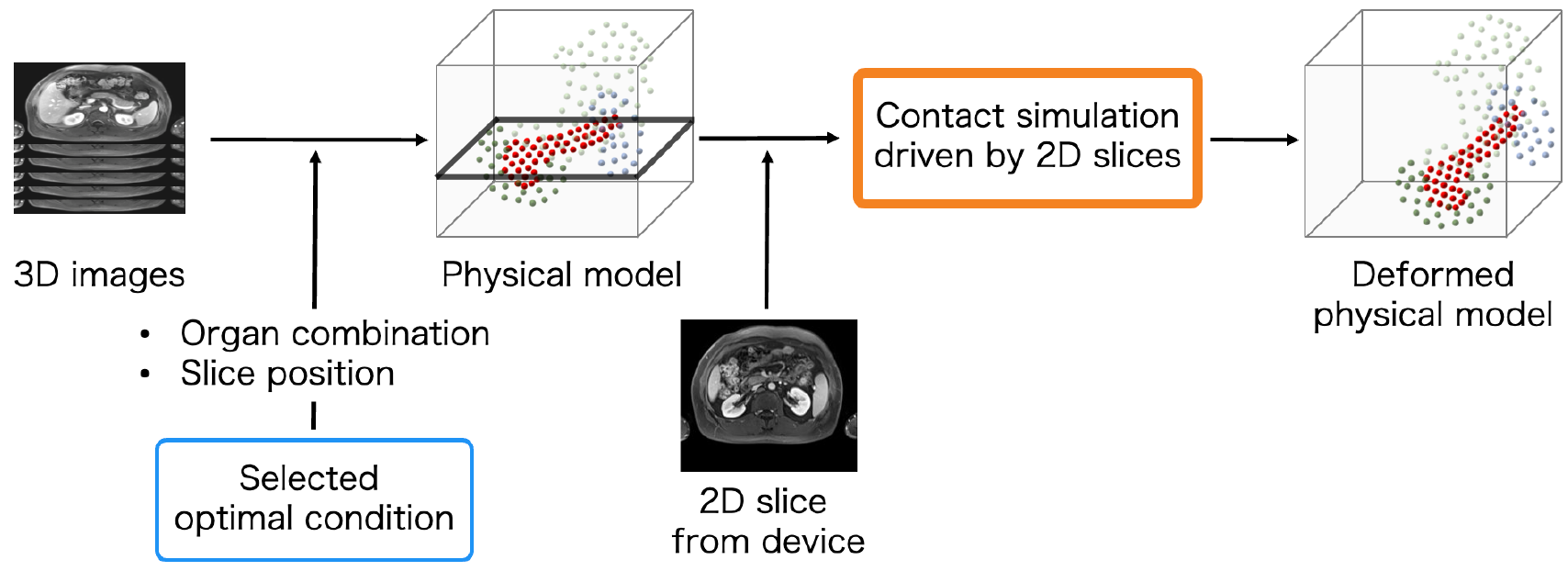}
  \caption{Conceptual diagram of the proposed framework.}
  \label{fig:ponch}
\end{figure}
\par Fig. \ref{fig:ponch} shows the concept of the proposed framework. The physical model is created from 3D images obtained before treatment. At this time, the optimal organ combination and slice position are selected from the regression model. During treatment, contact simulation is performed to estimate organ motion under the optimal conditions, using a 2D slice from the therapeutic device. 
\par This is an initial study to confirm the feasibility of the proposed framework. In particular, we tested our hypothesis regarding the consideration of surrounding organs. For these purposes, we used the 3D abdominal MR dataset and verified the effectiveness of the proposed framework for pancreatic motion.
\begin{figure*}[tb]
  \centering
  \includegraphics[width=\linewidth]{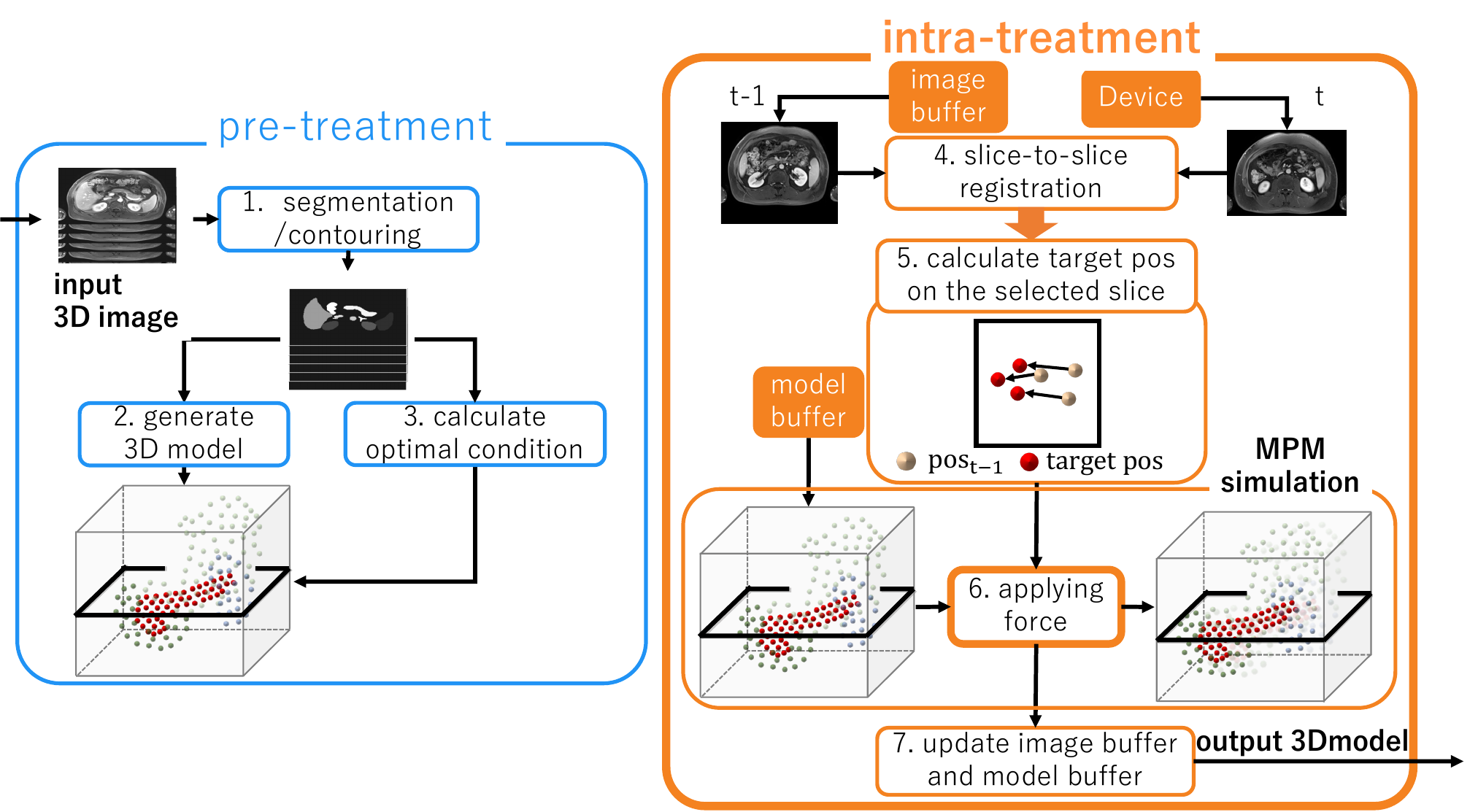}
  \caption{Flow of the proposed framework. The proposed framework has two steps. The first is the ``pre-treatment'' step. The physical model is generated and the optimal condition is selected. The second step is the ``intra-treatment'' step. By driving physical models based on 2D slices, the 3D motions of the organs are estimated.}
  \label{fig:concept}
\end{figure*}
First, to validate our hypothesis and confirm the basic performance of the proposed framework, the following three validations were conducted using the axial direction:
\begin{itemize}
    \item validation of the accuracy variation with combinations of organs under consideration
    \item comparison of the accuracy between the proposed framework and the volume-to-volume registration method, which is considered to be ideal
    \item validation of a regression model for selecting the optimal conditions.
\end{itemize}
Second, to validate the effectiveness of the proposed framework, we also conducted the following validation using the axial, sagittal, and coronal directions:
\begin{itemize}
    \item validation of the accuracy variation with combinations of slice directions.
\end{itemize}
In this study, we considered the stomach, duodenum, and left kidney, which are the surrounding organs close to the pancreas.

\section{METHODS}
\subsection{Overview of the 2D Slice-driven 3D Organ Motion Estimation Framework}
Fig.~\ref{fig:concept} illustrates the flow of the proposed framework.
 The proposed framework has two steps, the ``pre-treatment'' step and the ``intra-treatment'' step. In the ``pre-treatment'' step (section \ref{sec:pre_treatment}), 
\begin{enumerate}
    \item 3D physical models are generated from 3D data before treatment
    \item the optimal slice position and the combination of organs are selected for driving the 3D physical models
    \item the selected slice position is registered as the position to be captured by the MR-Linac.
    \item the position of the physical model corresponding to the slice position is registered.
\end{enumerate} 
In the ``intra-treatment'' step (section \ref{sec:intra-treatment}),
\begin{enumerate}
    \item deformation at that slice position is calculated from the image obtained from the device and the previous step
    \item the target position at the registered position of the physical model is calculated based on the deformation
    \item the entire 3D physical model is driven by applying a force so that the registered position of the physical model approaches the target position.
\end{enumerate}
Of these steps, No.2 of the ``pre-treatment'' and No.3 of the ``intra-treatment'' step are the distinctive features of this study. 

For the physical simulation, we used the Material Point Method (MPM)~\cite{mpm_book,mpm_snow}, which can efficiently handle the contact between different objects.
\subsection{Material Point Method}
MPM is a mesh-free method. The particles discretize the object, and the discrete equations are solved using a grid~\cite{mpm_book,mpm_snow}. The laws of conservation of mass and momentum are expressed as follows: 
\begin{align}
    \label{eqn:conservation_mass}
    \frac{D\rho}{D t} + \rho \nabla \cdot \bm{v}= 0, \\
    \label{eqn:conservation_momentum}
    \rho\frac{D \bm{v}}{D t} = \nabla \cdot \bm{\sigma} + \rho \bm{b},
\end{align}
where $\rho$, $\bm{v}$, $t$, $\bm{\sigma}$, and $\bm{b}$ denote density, velocity, time, stress, and body force, respectively, and $D/Dt$ denotes the material derivative.
The constitutive laws depend on the target of discretization.
\par There are three main processes in one loop. First, the physical properties are transmitted to the grid using the shape function.
Next, the discrete equation is solved on the grid. Finally, the particles are updated using the grid information and the grid is discarded.

\par We represent organs as physical models and run contact simulations between organs.
Therefore, a numerical method that can efficiently represent contact is desired.
Although many numerical methods require additional steps for handling contact, MPM can handle contact without any special processing, because the discrete equation is solved on a single grid.

\subsection{Pre-Treatment Step}
\label{sec:pre_treatment}
This step involves two processes that are required before treatment.
The proposed framework requires the segmentation of the organs to be used in advance to identify the organ regions in the image for these processes (process 1. in Fig.~\ref{fig:concept}).
\subsubsection{Model Generation (process 2. in Fig~\ref{fig:concept})}

In this process, the proposed framework generates a 3D MPM model from the 3D images captured before treatment. The therapeutic device captures only several 2D slices during treatment.
Since the proposed framework cannot generate 3D models in the ``intra-treatment'' step, the proposed framework generates them from pre-treatment 3D MRI or computed tomography (CT) images.
\subsubsection{Selection of Optimal Slice and Considered Organs (process 3. in Fig.~\ref{fig:concept})}
\label{sec:best_cs}

 In this process, the proposed framework determines the optimal slice position to drive the 3D MPM model, and the combination of discrete surrounding organs. The optimal condition determined by this step is used in the ``intra-treatment'' step. The thick black border in Fig.\ref{fig:concept} represents its selected slice position. 
\par The estimation accuracy is greatly affected by the conditions. A brute-force simulation would yield the optimal condition, but this would be extremely time-consuming. To select the optimal condition, the proposed framework requires an indicator used to estimate the accuracy of the proposed framework. We constructed a regression model to serve as an indicator.
When estimating 3D motion from 2D slices, the slice to be used is crucial. 
Organ combinations are also very important.
This is because considering all organs may not always be the optimal condition, as it depends on the positional relationship between the slices and the surrounding organs.
We therefore created a regression model that predicts the accuracy of the proposed framework based on the condition of the slice position and the combination of surrounding organs.
 The number of organ particles in each slice is used as input to regress the accuracy of the proposed method. 
Using this as an indicator, the proposed framework determines the optimal condition.
\par In order to train a regression model, a training data set must be created.
In this study, we first created a pseudo-deformed 3D dataset (section \ref{sec:GT}). Next, we created training data for the regression model from the results obtained by a brute force simulation on slice positions and organ combinations (section \ref{sec:various_organs}).

The regression model estimates the accuracy for all types of slice positions and organ combinations. In this study, the proposed framework used the stomach, duodenum, and left kidney as the surrounding organs, but the model is not limited to only these surrounding organs. \relax
 \relax
The slice position and the combination of surrounding organs in the case with the highest accuracy is the optimal condition. \relax
The proposed framework uses slice position and organ combinations at the time of the condition. In Fig.~\ref{fig:concept}, the black frame in the 3D model indicates a selected slice position. Using the information from the slice, the proposed framework estimates the 3D motion. \relax
Section~\ref{sec:various_organs} describes the effects of organ combinations. \relax
 \relax
 \relax
\subsection{Intra-Treatment Step}
\label{sec:intra-treatment}
In this step, the framework estimates 3D motion from 2D slices. The objective of this step is to generate 3D models at step $t$ by deforming 3D models at step $t-1$ based on images captured by the device at step $t$.

\subsubsection{Slice-to-Slice Registration to Obtain Motion Information (process 4. in Fig~\ref{fig:concept})}
\label{sec:slice-to-slice}
This process takes the optimal slice position obtained in the previous step as an input, and  computes the displacement of particles on the selected slice using the slice-to-slice registration between the stored image at step $t-1$ and the captured image at step $t$.  In this study, these images are masked by contoured images, but are not necessarily required. The parameters obtained from this registration are passed to the next process in order to calculate the target position of the particle on the selected slice in the physical model. 
\par We assume an affine transformation for image deformation.
Affine transformation is an image deformation method that handles four types of motions (rotation, scaling, shearing, and translation).
An affine transformation is given by:
\begin{equation}
    \label{eqn:affine_base}
    \bm{\tilde{y}} = A \left(\bm{\tilde{x}} - \bm{\tilde{c}}\right) + \bm{\tilde{c}} + \bm{\tilde{t}},
\end{equation}
where $A$ denotes the rotation/scaling/shearing transformation matrix, $\bm{x}$ denotes the fixed position, $\bm{y}$ denotes the moving position, $\bm{c}$ denotes the center of rotation, and $\bm{t}$ denotes the translation vector. Note that this equation uses the simultaneous coordinates ($\bm{\tilde{x}}, \bm{\tilde{y}}, \bm{\tilde{c}},  \bm{\tilde{t}}$).
Using this equation, the moving position $\bm{y}$ is calculated from the fixed position $\bm{x}$.
The proposed framework uses the matrix $A$ and vector $\bm{t}$ in the next process  (see Section~\ref{sec:drive}).

\subsubsection{Contact Simulation Based on the Deformation Information (process 5, 6. in Fig~\ref{fig:concept})}
This process drives 3D models using the affine transformation parameters calculated in Section \ref{sec:slice-to-slice}.
\par First, the target positions of the particles $\bm{x}_{i}^{t}, i \in \Omega_{\mathrm{cs}}$ on the selected slice $\Omega_{\mathrm{cs}}$ are calculated from the current positions $\bm{x}_{i}^{t-1}, i \in \Omega_{\mathrm{cs}}$ using

\label{sec:drive}
\begin{equation}
\label{eqn:affine transform}
     \bm{\tilde{x}}_{i}^{t} = A \left(\bm{\tilde{x}}_{i}^{t-1} - \bm{\tilde{c}}\right) + \bm{\tilde{c}} + \bm{\tilde{t}}.
\end{equation}
Next, the proposed framework performs a contact simulation between the pancreas and surrounding organs.
The proposed framework drives the 3D models by applying forces to the particles on the selected slice, and brings the particles closer to the target positions.

It then calculates the driving forces from the current positions $\bm{x}^{t-1}$ and target positions $\bm{x}^{t}$  of the particles using a proportional\-integral\-differential (PID) controller. The PID controller calculates the driving force $\bm{f}_{\mathrm{dri},i}$ so that the difference between the current and target values, i.e. $\bm{x}_{\mathrm{dif},i} =  \bm{x}_{i}^{t} - \bm{x}_{i}^{t-1}$,  becomes zero by manipulating proportional, differential, and integral elements as follows:
\begin{align}
    \label{eqn:drive}
    \begin{split}
    \bm{f}_{\mathrm{dri},i} &= K_{\mathrm{p}}  \bm{x}_{\mathrm{dif},i} \\
    & \quad + K_{\mathrm{d}} \frac{d\bm{x}_{\mathrm{dif},i}}{dt} + K_{\mathrm{i}}\int_0^{t} \bm{x}_{\mathrm{dif},i} dt,
    \end{split}
\end{align}
where $K_{\mathrm{p}}, K_{\mathrm{d}}$, and $K_{\mathrm{i}}$ denote proportional, differential, and integral gains, respectively.
The proposed framework drives particles on the selected slice using this force, and repeats the calculation until the relative error ($rel\_error$)
\begin{equation}
    \label{eqn:error}
    rel\_error = \sum_{i=0}^{N_\mathrm{cs}-1} \frac{\left| \bm{x}_{\mathrm{dif},i}\right|}{\left| \bm{x}_{i}^{t}\right|} 
\end{equation}
satisfies the convergence condition. In Equation~(\ref{eqn:error}), $N_\mathrm{cs}$ denotes the number of particles in $\Omega_{\mathrm{cs}}$.
\par The PID controller may oscillate or overshoot depending on parameters. Selecting parameters that neither oscillate nor overshoot is ideal, but difficult.
Therefore, we define convergence when the $rel\_error$ satisfies one of the following conditions:
\begin{itemize}
    \item less than the specified value (TH\_V) for the specified number of consecutive times (TH\_S),
    \item does not change up to the TH\_D decimal places for the specified number of consecutive times (TH\_S). Since the image space is compressed to $1\times1\times1$ in the simulation by MPM, the decimal point is taken into account.
\end{itemize}
Under these conditions, the proposed framework maintains the vibration within a certain range, whereas the value follows the target value. We set the $\mathrm{maximum\_force}$ to round the value above, in order to prevent the simulation from diverging because of the excessive driving force $\bm{f}_{\mathrm{dri}}$. Algorithm~\ref{algo:proposed} depicts the entire intra-treatment algorithm.
Variables with index $i$ are those of the grid and those with $p$ are those of particles. $\bm{F}$, $\bm{C}$, and $V$ denote deformation gradient, affine velocity, and volume, respectively. $\omega_{ip}=N_{\bm{i}}(\bm{x}_p)$ and $\nabla\omega_{ip}=\nabla N_{\bm{i}}(\bm{x}_p)$ were calculated from the interpolation function $N$ which connects particle $p$ and grid points $i$. 
$RD()$ in Algorithm~\ref{algo:proposed} is a function that rounds the input into the TH\_D digit, and $rel\_error_{bef}$ is a variable that stores $rel\_error$ of one previous step.
\par In this study, the proposed framework used an affine transformation. This could be replaced, however, by any method that can be used to calculate the target position. The use of nonlinear deformation, such as free-form deformation and machine-learning methods, as mentioned in Section~\ref{sec:intro}, may improve the accuracy of the proposed framework.

\begin{algorithm}

    \caption{Algorithm of intra-treatment}
    \label{algo:proposed}
    \centering
    \begin{algorithmic}
        
         \STATE Set up the Cartesian Grid: $\bm{x}_i^{0},\bm{v}_i^{0},m_i^{0}$
         \STATE Set step $n \leftarrow 0$
         \STATE Set up particle parameter:\\$\bm{x}_p^{0},\bm{v}_p^{0},\bm{\sigma}_p^{0},\bm{F}_p^{0},\bm{C}_p^{0},V_p^{0},m_p^{0},\rho_p^{0}$\\
        \STATE Set up parameters for convergence:\\ $cnt\_v,cnt\_d \gets 0, rel\_error, rel\_error_{\mathit{bef}} \gets \infty$
        \WHILE{}
            \FOR{each grid node \textit{i}}
                \STATE Reset all parameter: \\$m_{i}^n\leftarrow 0,\bm{v}_i^{n} \leftarrow \bm{0},\bm{f}_i^n \leftarrow \bm{0}$ \\
            
                \STATE Compute mass: $m_i^{n} \leftarrow \sum_{p} m_p \omega_{ip}^{n}$ \\
                \STATE Compute velocity: \\ $\bm{v}_i^{n} \leftarrow \frac{1}{m_i^{n}} \sum_{p} m_{p}\omega_{ip}^{n}(\bm{v}_p^{n} + \bm{C}_p^{n}(\bm{x}_i^{n} - \bm{x}_p^{n}))$
                \FOR{$\textbf{each particle } p \in \Omega_{\mathrm{cs}}$}
                    \STATE Compute driving~force ($f_{\textrm{dri},p}$) using Equation~\ref{eqn:drive}
                \ENDFOR
                \STATE Compute force: \\$\bm{f}_i \leftarrow  \sum_{p} \left(-V_p^{n}\bm{\sigma}_p\nabla \omega_{ip}^{n} +  V_p^n \bm{f}_{\textrm{dri},p} \omega_{ip}^{n}\right)$
                \STATE Compute pseudo velocity: \\$\bm{v}_i^{*} \leftarrow \bm{v}_i^{n} + \frac{\bm{f}_i^{n}}{m_i}\Delta t$
                \STATE Reflect boundary condition and compute true velocity $\bm{v}_i^{n+1}$ from $\bm{v}_i^{*}$
            \ENDFOR
            \FOR{$\textbf{each particle } p$}
                \STATE Compute velocity: $\bm{v}_{p}^{n+1} \leftarrow \sum_{i} \omega_{ip}^{n}  \bm{v}_i^{n+1}$
                \STATE Update position: $\bm{x}_p^{n+1} \leftarrow \bm{x}_p^{n} + \Delta t\bm{v}_p^{n+1}$
                \STATE Compute and damp affine velocity: $\bm{C}_p^{n+1}$
                \STATE Compute deformation gradient: \\ $\bm{F}_p^{n+1} \leftarrow \left(\bm{I} +  \Delta t \sum_{i} \nabla \omega_{ip}^{n} \bm{v}_i^{n+1} \right)\bm{F}_p^{n}$
                \STATE Singular value decomposition: $\bm{U}, \bm{\Sigma}, \bm{V} \leftarrow \bm{F}_p^{n+1}$
                \STATE Compute stress: $\bm{\sigma}_p \leftarrow 2 \mu \left(\bm{F}_p^{n+1} - \bm{U}\bm{V}^{\textrm{T}}\right)\bm{F}_{p}^{\textrm{T},n+1} + \lambda \left(\det \bm{\Sigma} - 1\right)\det \bm{\Sigma} \bm{I}$
            \ENDFOR
            
            \STATE Compute  $rel\_error$ using Equation~\ref{eqn:error} \IF{$\mathrm{max}\left(cnt\_v, cnt\_d\right) \geq  \mathrm{TH\_S}$}
                \STATE $\textbf{break}$
            \ELSE
                \IF{$rel\_error \leq \mathrm{TH\_V}$ }
                    \STATE $cnt\_v \gets cnt\_v + 1 $
                \ELSE
                    \STATE $cnt\_v \gets 0$
                \ENDIF
                \IF{$RD(rel\_error) == RD(rel\_error_{\mathrm{bef}})$ }
                    \STATE $cnt\_d \gets cnt\_d + 1 $
                \ELSE
                    \STATE $cnt\_d \gets 0$
                \ENDIF
            \ENDIF
            \STATE $rel\_error_{\mathrm{bef}} \gets rel\_error$
         \ENDWHILE
    \end{algorithmic}
\end{algorithm}

\section{EXPERIMENTAL SETUP}
\label{sec:data_settings}
\subsection{Data Specification and Preprocessing}
\label{sec:data_info}
The effectiveness of the proposed framework was verified by generating a pseudo-test dataset using a 3D MRI dataset.
We used 3D MR images from the Multi-Modality Abdominal Multi-Organ Segmentation Challenge 2022~\cite{amos}, where 
all data had been approved by the Research Ethics Committees of Longgang District People’s Hospital (Shenzhen, China) (reference number: 2021077) and Longgang District Central Hospital (Shenzhen, China) (reference number: 2021ECJ012).
This challenge distributes abdominal CT and MR images collected from multi-center, multi-vendor, multi-modality, multi-phase, and multi-disease patients, each with voxel-level annotations of 15 abdominal organs.
We extracted the data with the constraint that no more than 40 slices with a pancreas could be present, and used 20 images out of the 40 3D MR images provided as training sets.
We considered the stomach, duodenum, and left kidney as surrounding organs to the pancreas.
For process 1. in Fig~\ref{fig:concept}, we used the following two processes to pre-process the data.
\begin{itemize}
    \item shaping each image into a square image by zero padding
    \item extracting only the organs in use from MR images using the contoured images as the mask images
\end{itemize}
Table~\ref{tbl:data_parameters} shows the parameters of the dataset.

\begin{table}
 \caption{Parameters of dataset} \label{tbl:data_parameters}
  \centering
  \begin{tabular}{c||c}
  Parameter & Value \\
  \hline \hline
  Num. of dataset & 20 \\ \hline 
  Num. of motion & 8 \\ \hline
  Pixel spacing & $1.22 \pm 0.245$ mm \\ \hline
  Slice thickness & $2.93 \pm 0.238$ mm\\ \hline
  Image size(x,y) & $192 \sim 576$ \\ \hline
  Image size(z) & $64 \sim 100$ \\ \hline
 \end{tabular}
\end{table}

\par MR-Linac is capable of capturing images in three directions (axial, sagittal, and coronal). 
Therefore, we performed three basic validations of the proposed framework using one direction, followed by validation using slices in multiple directions.
For the first three validations, we used the axial direction, which is the slice direction of the dataset.

\subsection{Generation of the Ground Truth}
\label{sec:GT}
Two points were considered when generating pseudo-test data for the ground truth (GT).
The first was the `` deformation direction.'' As mentioned in Section~\ref{sec:data_info}, we used the axial direction for the first three validations. 
For the three validations using a slice, we needed to generate the GT so that axial motion was dominant, but motions in other directions were also present.
The second was the ``expression of contact between organs.'' Because this study aimed to estimate the aperiodic motion caused by organ-to-organ contact, we needed to give the GT the motion caused by organ-to-organ contact.
We used the following four steps to generate a GT that took the above two points into consideration:
\begin{enumerate}
    \item construction of a 2D transformation matrix from the rotation, shearing, and scaling parameters.
    \item deformation of all slices using the transformation matrix made in  step 1.
    \item calculation of the target positions of all particles in the 3D models
    \item running of the simulation in the same way as Section~\ref{sec:drive}. By contrast, the $\bm{f}_{\mathrm{dri}}$ is applied to all particles and $rel\_error$ is calculated using all particles.
\end{enumerate}
\par We calculated the transformation matrix $A$ as the product of the rotation matrix $R$, the shearing matrix $G$, and the scaling matrix $S$,
\begin{align}
\label{eqn:RGS}
    \nonumber
    R &= 
    \begin{pmatrix}
    \cos \left(r/2\pi\right) & -\sin \left(r/2\pi\right) & 0 \\
    \sin \left(r/2\pi\right) & \cos \left(r/2\pi\right) & 0 \\
    0 & 0 & 1
    \end{pmatrix}, \\
    \nonumber
    G &= 
    \begin{pmatrix}
    1 & \tan \left(g/2\pi\right) & 0 \\
    \tan \left(g/2\pi\right) & 1 & 0 \\
    0 & 0 & 1
    \end{pmatrix}, \\
    \nonumber
    S &= 
    \begin{pmatrix}
    s & 0 & 0 \\
    0 & s & 0 \\
    0 & 0 & 1
    \end{pmatrix}, \\ 
    A &= RGS,
\end{align}
where $r, g, s$ are the rotation, shearing, and scaling parameters, respectively.
\par
\begin{figure*}
    \includegraphics[width=\linewidth]{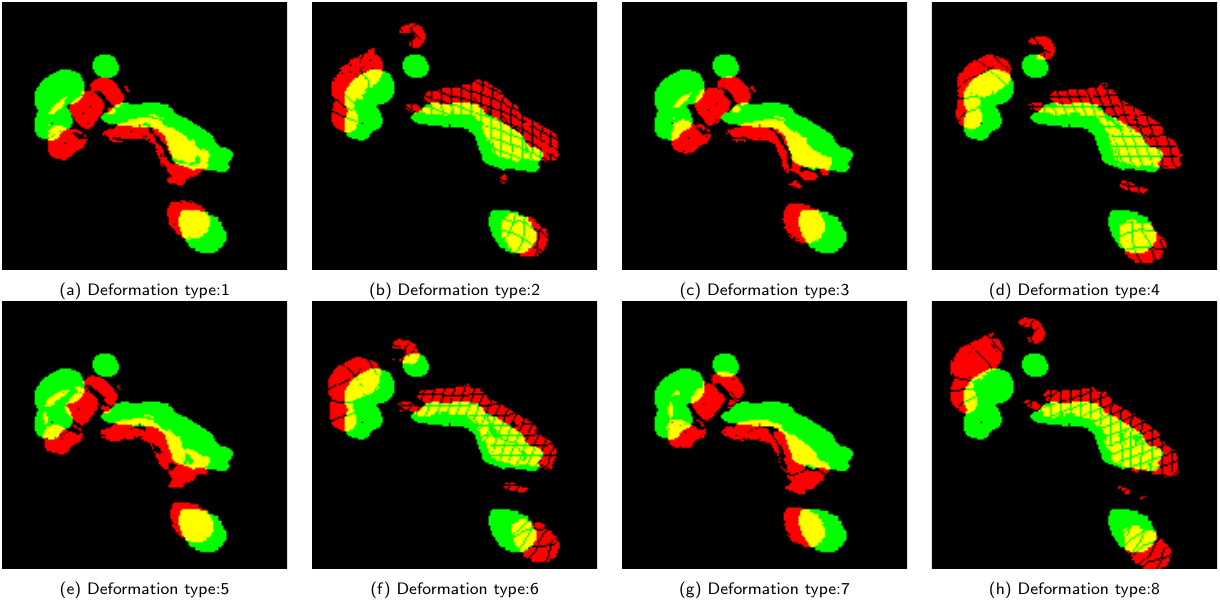}
\caption{Examples of each deformation. The green, red, and yellow colors represent initial regions, deformed regions, and the regions where both initial and deformed regions overlap, respectively. Only the simulated organs are extracted in advance using labels.}
\label{fig:motion_example}
\end{figure*}

 To evaluate the accuracy for aperiodic motion, we generated the GT using all possible combinations of these parameters ($r = \{-2.0, 2.0\}, g = \{-2.0, 2.0\}, s = \{0.8, 1.2\}$) for each of the 20 data. The number of deformations was $2^3 = 8$; therefore, the number of GT was $20 \times 8 = 160$.
Table~\ref{tbl:motion_types} lists the combinations of parameters used to generate the GT, and Fig.~\ref{fig:motion_example} shows examples of images with the deformation applied. The green, red, and yellow colors represent the initial regions, deformed regions, and regions where the initial and deformed regions overlap, respectively. In Fig.~\ref{fig:motion_example}, only the simulated organs were extracted in advance using labels. Fig.~\ref{fig:motion_example} shows that each organ is deformed to different degrees in the direction of the image center, as well as outward.

\begin{table}
 \caption{Parameters of deformation types} \label{tbl:motion_types}
  \centering
  \begin{tabular}{c||c|c|c}
    Deformation type & R & G & S \\ \hline\hline
    1 & -2.0 & -2.0 & 0.8\\ \hline
    2 & -2.0 & -2.0 & 1.2 \\ \hline
    3 & -2.0 & 2.0 & 0.8\\ \hline
    4 & -2.0 & 2.0 & 1.2 \\ \hline
    5 & 2.0 & -2.0 & 0.8\\ \hline
    6 & 2.0 & -2.0 & 1.2 \\ \hline
    7 & 2.0 & 2.0 & 0.8\\ \hline
    8 & 2.0 & 2.0 & 1.2 \\ \hline
 \end{tabular}
\end{table}

\subsection{Configuration of the Framework}
The ``pre-treatment'' step generates the MPM model from the pre-processed data, as described in Section~\ref{sec:data_info}.
This step assigns a pixel to a particle and compresses the image space by $1\times 1\times 1$. For the constitutive law, we used the energy density function proposed by Stomakhin et al.~\cite{Stomakhin}.

\par The outputs of this framework depend largely on the accuracy of the slice-to-slice registration in Section~\ref{sec:slice-to-slice}. The purpose of this study was to verify the effectiveness of interpolating out-of-plane motion using a physical simulation.
To avoid the effect of the error caused by slice-to-slice registration, the ``intra-treatment'' step uses the affine parameters to generate the GT, as described in Section~\ref{sec:GT}, instead of ``slice-to-slice registration (process 4. in Fig~\ref{fig:concept}).''
Table~\ref{tbl:sim_parameters} lists the parameters used in the simulations. We determine the PID parameters empirically.
\begin{table}
 \caption{Parameters of simulation} \label{tbl:sim_parameters}
  \centering
  \begin{tabular}{c||c|c}
  & Parameter & Value \\
  \hline \hline
  & Num. of particles & $7160\sim71679$ \\ \cline{2-3}
  Pancreas~\cite{pancreas} & $\lambda$ & $5.76\times 10^4$ Pa \\ \cline{2-3}
  & $\mu$ & $1.17\times 10^3$ Pa \\ \hline 
  & Num. of particles & $6914 \sim 130371$ \\ \cline{2-3}
  Stomach~\cite{stomach_kidney} & $\lambda$ & $8.32 \times 10^7$ Pa \\ \cline{2-3}
  & $\mu$ & $1.68 \times 10^5$ Pa \\ \hline 
  & Num. of particles & $5177 \sim 30234$ \\ \cline{2-3}
  Duodenum~\cite{duodenum} & $\lambda$ & $8.11\times 10^7$ Pa \\ \cline{2-3}
  & $\mu$ & $3.38 \times 10^6$ Pa \\ \hline   
  & Num. of particles & $21197 \sim 125083$ \\ \cline{2-3}
  Left kidney~\cite{stomach_kidney} & $\lambda$ & $3.99\times 10^6$ Pa \\ \cline{2-3}
  & $\mu$ & $8.01\times 10^3$ Pa \\ \hline   
  & Time steps & $4.25 \times 10^{-7}$ \\ \cline{2-3}
  & TH\_V & $0.0015$ \\ \cline{2-3}
  & TH\_S & $100$ \\ \cline{2-3}
  Simulation & TH\_D & $4$ \\ \cline{2-3}
  & Maximum\_force & $1.00 \times 10^{11}$ \\ \cline{2-3}
  & $K_{\mathrm{prop}}$ & $1.00 \times 10^{8}$ \\ \cline{2-3}
  & $K_{\mathrm{diff}}$ & $8.00 \times 10^{7}$ \\ \cline{2-3}
  & $K_{\mathrm{inte}}$ & $1.00 \times 10^{7}$ \\ \hline
 \end{tabular}
\end{table}

\subsection{Quantitative Indices}
Two metrics were used to evaluate the proposed framework. The first was the average error (mm) between the position of the GT ($\bm{x}_{\mathrm{GT}, i}$) and the estimated position ($\bm{x}_{\mathrm{est}, i}$)  of each pixel in $i, 1 \leq i \leq N_p$ ($N_p$ is the number of pancreatic particles), expressed by 
\begin{equation}
    \label{eqn:metric_error}
    error = \frac{1}{N_p} \sum_{i=1}^{N_p}\|\bm{x}_{\mathrm{GT},i} - \bm{x}_{\mathrm{est},i}\|.
\end{equation}
We refer to this metric as $error$ below.
The second was the Dice score, which is expressed by 
\begin{equation}
    \label{eqn:metric_dice}
    dice = \frac{2 |\Omega_{\mathrm{GT}} \cap \Omega_{\mathrm{est}}| }{|\Omega_{\mathrm{GT}}| + | \Omega_{\mathrm{est}} |},
\end{equation}
calculated from the pancreatic region of the GT ($\Omega_{\mathrm{GT}}$) and the estimation ($\Omega_{\mathrm{est}}$).
Hereafter, we refer to this metric as $dice$.

\section{EXPERIMENTAL RESULTS AND DISCUSSION FOR ONE SLICE}
\label{sec:result1}
For validation, we ran a brute-force simulation with respect to slice positions, organ combinations, and deformation types for 20 volumes of data.
For the validation in Section~\ref{sec:various_organs} and Section~\ref{sec:vs_elastix}, we took the best result among all the slices as the result of the proposed framework in every 1280 data (20 MR images $\times$ 8 deformations $\times$ 8 combinations).
The simulation environment was Amazon EC2 c5a.24xlarge (3.3 GHz, 96 cores) (Amazon.com, Inc., Seattle, Washington, U.S.), and we ran simulations of approximately 90 cases concurrently for each measurement.

\subsection{Validation of the Effect of Considering Contacts on Accuracy}
\label{sec:various_organs}
\begin{figure*}[tb]
  \centering
  \includegraphics[width=\linewidth]{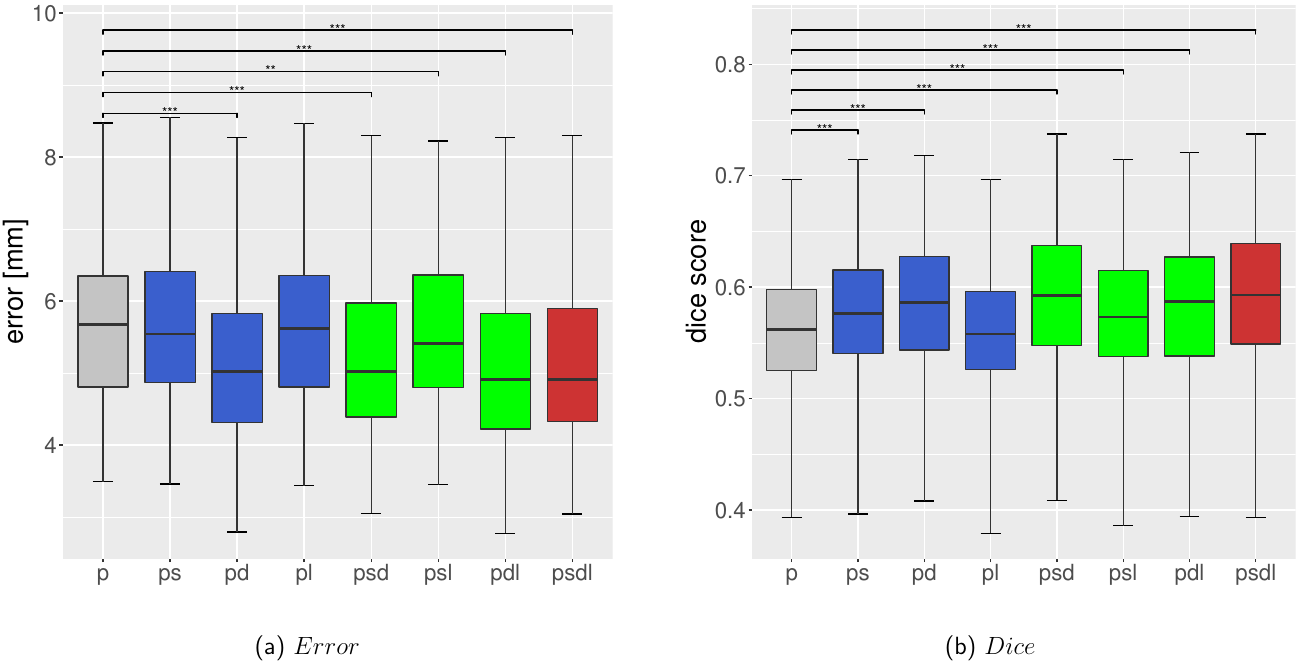}
  \caption{Box plots of each metric. The letters of p, s, d, and l indicate the pancreas, stomach, duodenum, and left kidney, respectively. These graphs are color-coded according to the number of organs used (gray: zero, blue: one, green: two, red: three). This only shows significant differences from multiple comparisons between ``pancreas (p)'' and other combinations (* $<$ 0.05, ** $<$ 0.01, *** $<$ 0.001).}
  \label{fig:box_contact}
\end{figure*}

Fig.~\ref{fig:box_contact} shows the results of $error$ and $dice$ for each organ combination.
For the sake of simplicity, as shown in Fig.~\ref{fig:box_contact} only shows significant differences from multiple comparisons between ``pancreas (p)'' and the other combinations.
Table~\ref{tbl:metric_vals} lists the means and standard deviations of $error$ and $dice$ for the initial condition and each combination of simulated organs.
\begin{table}
 \caption{Metric values for all conditions. The letters of p, s, d, and l represent the pancreas, stomach, duodenum, and left kidney, respectively.} 
 \label{tbl:metric_vals}
  \centering
  \begin{tabular}{c||c|c}
  Metric & $Error$ (mm) & $Dice$ \\
  \hline \hline
  Initial & $9.11\pm1.40$ & $0.408 \pm 0.0861$ \\ \hline
  p&$5.64\pm1.09$&$0.559\pm0.0587$ \\ \hline
  ps&$5.64\pm1.06$&$0.575\pm0.0554$ \\ \hline
  pd&$5.09\pm1.09$&$0.585\pm0.0611$ \\ \hline
  pl&$5.59\pm1.10$&$0.557\pm0.0581$ \\ \hline
  psd&$5.19\pm1.08$&$0.591\pm0.0615$ \\ \hline
  psl&$5.54\pm1.08$&$0.575\pm0.0568$ \\ \hline
  pdl&$5.06\pm1.14$&$0.583\pm0.0619$ \\ \hline
  psdl&$5.11\pm1.09$&$0.591\pm0.0631$ \\ \hline
 \end{tabular}
\end{table}
Almost all combinations showed a significant improvement in accuracy, whereas no significant difference was found for the conditions: ``pancreas + stomach (ps)'' and ``pancreas + left kidney (pl)'' for $error$ and the condition: ``pancreas + left kidney (pl)'' for $dice$.
This confirmed the effectiveness of considering not only the target organ (pancreas) but also the surrounding ones.
\par
\begin{figure}[tb]
  \centering
  \includegraphics[width=\linewidth]{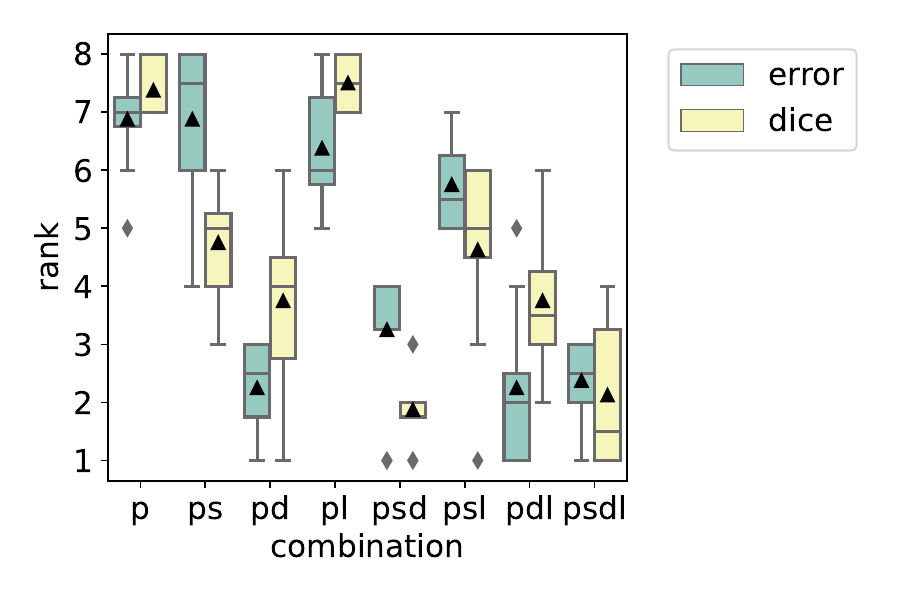}
  \caption{Ranks of each deformation. Each black triangle represents the mean.}
  \label{fig:rank}
\end{figure}

 Next,  to investigate the effects of organ combinations with respect to the type of deformity, we ranked the organ combinations by accuracy  for each of the types. Because there were eight combinations, the best rank was denoted ``1,'' and the worst ``8.''
Fig.~\ref{fig:rank} presents the ranking results.
In Fig.~\ref{fig:rank}, the number of organs considered and the ranking of accuracy did not show similar trends.
For example, the rank was particularly high when the duodenum was included, and the rank of ``psdl'' was not necessarily the highest. 
These results confirm that organ type is important to consider when aiming to improve accuracy.
These results support the importance of selecting the optimal combination of organs in the proposed framework, and the need for regression models.
\par In terms of simulation runtime, the longest was $2587 \pm1263$ s (approximately 43 minutes) when all organs were considered, and the shortest was $844 \pm 269$ s (approximately 15 minutes) when only the pancreas was considered. Because we ran the simulation in approximately 90 cases concurrently, these were not pure runtimes. Radiotherapy takes approximately several tens of minutes, and an MR-Linac is ideally able to acquire slices in real time. Therefore, to estimate the motion of organs using a larger number of images during treatment, it is necessary to accelerate these computations.
\par There are two major speed-up points in the proposed framework.
The first is ``the number of loops until convergence.'' As shown in Table~\ref{tbl:sim_parameters}, the proposed framework requires 100 steps to judge convergence (TH\_S). Consequently, the proposed framework requires hundreds or thousands of loops until convergence.
In this study, we used the same PID parameters for all of the data, since these had been empirically determined previously.
Therefore, by using the maximum likelihood parameters for individual data, the effects of overshoot and oscillation can be reduced and the framework can reduce TH\_S.
The second is ``the simulation time of one loop.'' The mean one-loop time under the ``psdl'' condition was $3.18 \pm 1.31$ s, and the mean time under the ``p'' condition was $1.16 \pm 0.212$ s. We assigned a pixel in the images to a particle in the physical models, and the framework did not use GPU.
By optimizing the mapping of the images to physical models, and by taking advantage of the GPU~\cite{mpm_gpu}, the proposed framework can be made to run much faster.

\subsection{Comparison Between the Proposed Framework and Volume-to-Volume Registration}
\label{sec:vs_elastix}
 Since the therapeutic device can capture only a few slices, it is not possible to use volume-to-volume registration.
However, we were able to use the 3D images of the pseudo-dataset that we created, mentioned in section \ref{sec:GT}. Therefore, in this section, we used volume-to-volume registration as the comparison method to set the accuracy that the proposed framework aimed for. 
For volume-to-volume registration, we used an affine transformation by elastix~\cite{elastix}.
Elastix is a medical image registration toolbox that can perform image registration using various types of deformations and metrics for both rigid and nonrigid objects. 
\par Table~\ref{tbl:metric_vals} showed that all combinations of the proposed framework improved the results compared to the initial condition, which did not involve estimation. The most accurate condition of the proposed framework ($error$: pdl, $dice$: psd) reduced $error$ by $45\%$ and increased $dice$ by $45\%$.
In terms of volume-to-volume registration, the accuracy was $3.34 \pm 1.46$ for $error$ and $0.646 \pm 0.114$ in $dice$, according to elastix.
All combinations of the proposed framework were inferior to the volume-to-volume registration accuracy for both the metrics.
\par By contrast, there were several cases where the proposed framework showed better results.
In the $error$ metric of ``psdl,'' $14.4\%$ of the 160 cases had better results than volume-to-volume registration.
In the $dice$ metric of ``psdl,'' $26.9\%$ of the 160 cases had better results than the volume-to-volume registration, and $33.8\%$ of the 160 cases had better results than the median of volume-to-volume registration. 
Therefore, it was suggested that the proposed framework was able to achieve motion estimation comparable to volume-to-volume registration, by interpolation using physical models from a single image.  Further comparisons using multiple slices are described in section \ref{sec:multi}. 
\par
\begin{figure*}[tb]
  \centering
  \includegraphics[width=\linewidth]{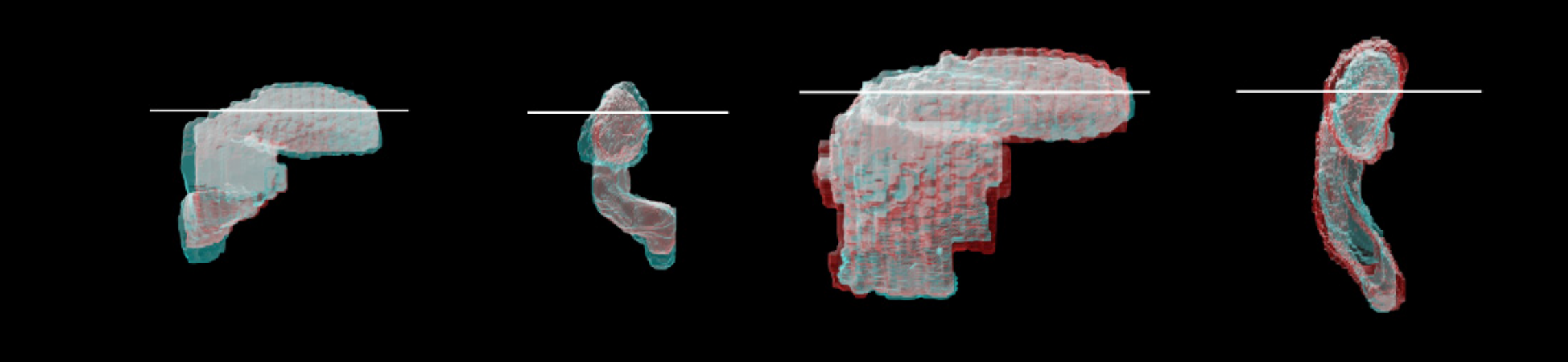}\\
  \hspace{10mm}
Best {\it error} : Front view{\hspace{7mm}} Best {\it error} : Side view {\hspace{7mm}} Worst {\it error} : Front view {\hspace{7mm}} Worst {\it error} : Side view \\
  \vspace{2mm}
  \includegraphics[width=\linewidth]{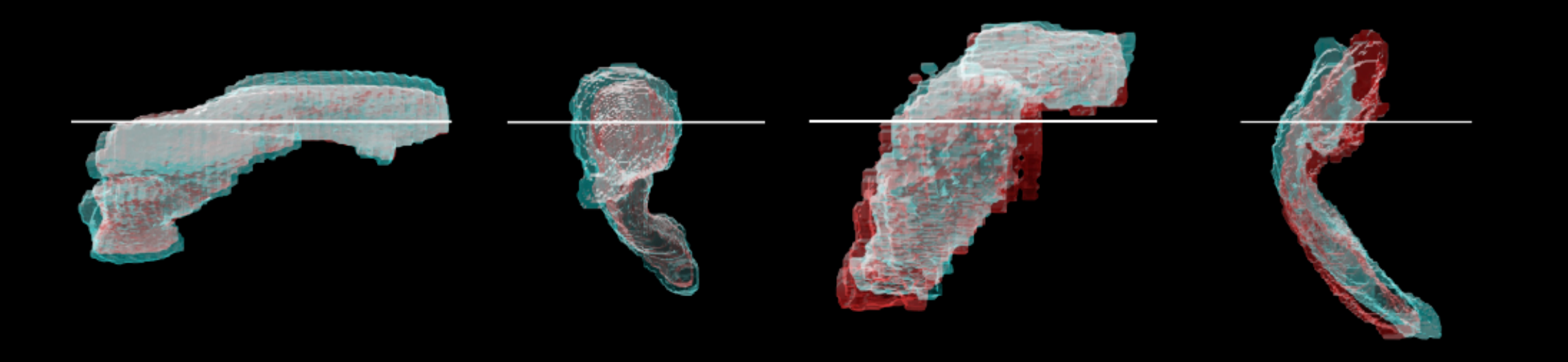}\\
  \hspace{10mm}
Best {\it dice} : Front view{\hspace{7mm}} Best {\it dice} : Side view {\hspace{7mm}} Worst {\it dice} : Front view {\hspace{7mm}} Worst {\it dice} : Side view \\
  \caption{Registration result of pancreas with $error$ (upper) and $dice$ (lower) in ``psdl.'' The best $error$ was $3.04$, the worst $error$ was $8.30$, the best $dice$ was $0.737$, and the worst $dice$ was $0.393$. The light blue region is the estimated region, while the red region is the GT region. The white line indicates the selected slice position.}
  \label{fig:qualitative}
\end{figure*}

 Fig.~\ref{fig:qualitative} shows the pancreatic regions when the best and worst values are marked in ``psdl'' by each metric.
The results for the best accuracy in Fig.~\ref{fig:qualitative} show that most of the estimated and GT regions overlapped, qualitatively confirming the accuracy of the estimation.
By contrast, it can be seen that there are areas of misalignment in the upper and lower parts of the image, and that these misalignments are very apparent in those with the worst accuracies. 
These misalignments lead to an accuracy gap between the proposed framework and volume-to-volume registration. In this validation, we used only one axial slice in the ``slice-to-slice registration (process 4. in Fig~\ref{fig:concept})'' part, as shown in Fig.~\ref{fig:concept}.
The further from the slice position to be driven, the greater the degree of freedom of motion, which is thought to lead to greater registration errors.

\subsection{Regression Model to Select Optimal Slice Position and Organ Combination}
\label{sec:cross_section}
In Section~\ref{sec:various_organs}, it was confirmed that the combination of organs to be modeled was a key parameter. In addition to this, it became obvious that the position of the driving slice was also crucial.
We therefore used ``slice position'' and ``the number of particles in each organ on the considered slice'' as explanatory variables and ``$error$ or $dice$ when the proposed framework is run under that condition'' as objective variables. Because the number of slices and the sizes of organs vary depending on the data, we standardized the explanatory variables. To make the explanatory and objective variables correspond one-to-one, we used the mean value with eight deformations of $error$ or $dice$ when the proposed framework is run under that condition.
The number of generated datasets was 4552, and we divided the dataset using an 8:2 ratio between the training set (3641) and test set (911).
We used eXtreme Gradient Boosting (xgb)~\cite{xgb} for regression, and performed hyperparameter tuning with 5-fold cross-validation using Optuna~\cite{optuna}.
Tuning was stopped when the maximum R2 score was not updated 100 times.
Table~\ref{tbl:cs_param_output} lists the R2 score, RMSE, and feature importance (FI) of each metric.

\begin{table}
 \caption{Best result for each model. FI denotes feature importance.} \label{tbl:cs_param_output}
\centering
\scalebox{1}{
  \begin{tabular}{c||l|c}
  & Parameter & Value \\
  \hline \hline
        & Train r2 score & 0.992\\ \cline{2-3} 
        & Test r2 score & 0.936 \\ \cline{2-3}
        & Train RMSE & 0.129 \\ \cline{2-3} 
        & Test RMSE & 0.357 \\ \cline{2-3}
$Error$ & FI : cs & 0.207 \\ \cline{2-3}
        & FI : pancreas & 0.371 \\ \cline{2-3}
        & FI : stomach & 0.191 \\ \cline{2-3}
        & FI : duodenum & 0.119 \\ \cline{2-3}
        & FI : left kidney & 0.112 \\ \hline
        & Train r2 score & 0.960 \\ \cline{2-3} 
        & Test r2 score & 0.903 \\ \cline{2-3}
        & Train RMSE & 0.0134 \\ \cline{2-3} 
        & Test RMSE & 0.0202 \\ \cline{2-3}
 $Dice$ & FI : cs & 0.183 \\ \cline{2-3}
        & FI : pancreas & 0.503 \\ \cline{2-3}
        & FI : stomach & 0.156 \\ \cline{2-3}
        & FI : duodenum & 0.0867 \\ \cline{2-3}
        & FI : left kidney & 0.0769 \\ \hline
 \end{tabular}
 }
\end{table}

Regarding $error$ for the test data, the R2 score was $0.936$ and the RMSE was $0.357$ mm. The mean of all of the data obtained from a brute-force run was $5.36 \pm 1.12$ mm, the minimum was $2.77$ mm, and the maximum was $8.55$ mm.
Regarding $dice$ for the test data, the R2 score was $0.903$ and the RMSE was $0.0202$. The mean of all of the data obtained from a brute-force run was $0.577 \pm 0.0608$, the minimum was $0.379$, and the maximum was $0.737$. From these results, we concluded that these regression models were able to calculate the accuracy of the proposed framework under every condition. Using these models, the framework can determine optimal slice position and organ combinations in the ``calculate optimal condition (process 3. in Fig.~\ref{fig:concept})''.
\par The feature importance shows that not only the slice position, but also the surrounding organs, contribute to the regression models. It has been suggested that the combination of organs is important, and these results were in agreement with those from Section~\ref{sec:various_organs}. The mean computation time of the ``calculate optimal condition (process 3. in Fig~\ref{fig:concept})'' using these regression models was $1.48 \pm 0.151$ ms (number of slices: $20 \sim 35$, number of surrounding organs: $3$). 
\par In the above three verifications, the axial direction, which is the slice direction of the data set, was used to confirm the feasibility of the proposed framework as an initial study. However, in clinical MR-guided radiation therapy, motion management is performed using sagittal or coronal slices. Therefore, in order to verify the practicality, it is necessary to apply sagittal slices or coronal slices instead of axial slices as the primary slice orientation.

\section{EXPERIMENTAL RESULTS AND DISCUSSION FOR MULTIPLE SLICES}
\label{sec:multi}
In Section~\ref{sec:vs_elastix}, we described the registration error of the region that exists at a certain distance away from the selected slice. It was also suggested that the proposed framework has power comparable to volume-to-volume registration.
Because the therapeutic device can capture both sagittal and coronal slices during treatment, the accuracy of the proposed framework can be improved by using multiple directions.
In this section, we verify the accuracy using a combination of slice directions.
\subsubsection{Validation Settings}
 In the eight organ combinations, we used ``psdl,'' because ``psdl'' showed good results and little variation in the rankings between the metrics, as is shown in Fig.~\ref{fig:rank}. The dataset generated in Section~\ref{sec:GT} had dominant axial motion, but was deformed in other directions due to the contact simulation. Therefore, we used the same 160 dataset (20 MR images $\times$ 8 deformations) for validation in Section~\ref{sec:GT}.
\par In contrast to Section~\ref{sec:result1}, the following two points were considered.
The first was ``the target positions of particles on the selected slice (process. 5 in Fig.~\ref{fig:concept}).'' The affine parameters in Table~\ref{tbl:motion_types} were used to calculate the target position in Section~\ref{sec:result1}. By contrast, there was no transformation matrix in the sagittal and coronal directions.
Therefore, regarding ``slice-to-slice registration (process 4. in Fig.~\ref{fig:concept}),'' the framework used the correct position from the GT as the target position, to avoid affecting the slice-to-slice registration errors.
\par The second was the ``driving slice position in three directions.'' 
For the axial direction, we determined the optimal position when we ran the simulation using brute force. However, the optimal positions in the sagittal and coronal directions were still unknown. 
We assumed that there was a correlation between the number of particles in each organ and the accuracy, regardless of the direction.
Therefore, we created another xgb model to determine the conditions in the sagittal and coronal directions. 
We used the number of particles in each organ as the explanatory variable, and the $dice$ metric ranked within slices in each of the 160 cases as the objective variable.
Regression was performed in the same manner as detailed in Section~\ref{sec:cross_section}.
As the R2 score for the test data was $0.859$, we considered this regression model to be valid.
\par We performed the validation under the following four conditions:
\begin{itemize}
    \item axial only (a)
    \item axial + coronal (ac)
    \item axial + sagittal (as)
    \item axial + sagittal + coronal (asc).
\end{itemize}
For each sagittal and coronal direction, we used the top three slices output by the regression model as candidates for the optimal slice position, and considered the best data among them as the result.
\subsubsection{Result}
\begin{figure*}[tb]
  \centering
  \includegraphics[width=\linewidth]{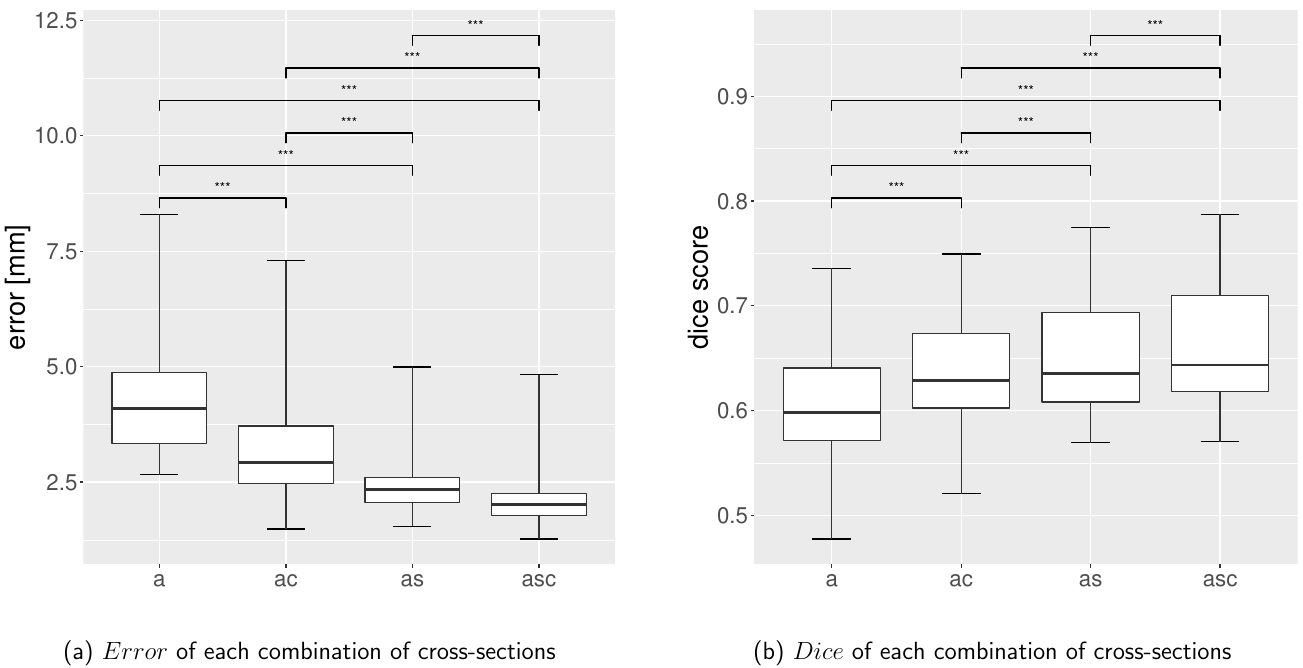}
  \caption{Box plots of each metric: a, c, and s represent the axial, coronal, and sagittal directions, respectively. (* $<$ 0.05, ** $<$ 0.01, *** $<$ 0.001)}
  \label{fig:multi_boxplot}
\end{figure*}

Because we used the GT's positions for the target positions on the selected slice, each metric was calculated without the particles on the selected slices.
Fig.~\ref{fig:multi_boxplot} shows the boxplot, and Table~\ref{tbl:multi_metric_vals} lists the means and standard deviations of each combination of the proposed framework.
\begin{table}
 \caption{Metric values under the tested conditions, which include axial, coronal, and sagittal orientations.} 
 \label{tbl:multi_metric_vals}
  \centering
  \begin{tabular}{c||c|c}
  Metric & $Error$ (mm) & $Dice$\\
  \hline \hline
  Axial only & $4.26 \pm 1.16$ & $ 0.610 \pm 0.0519$\\ \hline
  Axial + coronal & $3.12 \pm 0.989$ & $ 0.636 \pm 0.0488$\\ \hline
  Axial + sagittal & $2.42 \pm 0.584$ & $ 0.650 \pm 0.0490$\\ \hline
  Axial + sagittal + coronal & $2.13 \pm 0.598$ & $ 0.662 \pm 0.0504$\\ \hline
 \end{tabular}
\end{table}
Fig.~\ref{fig:multi_boxplot} and Table~\ref{tbl:multi_metric_vals} show that the accuracy of the proposed framework significantly increased as the number of slices used increased.
\par The sagittal direction was more accurate than the coronal direction for both metrics.
The maximum number of pancreatic particles in each slice of the 20 cases was $336 \pm 141$ in the coronal direction, and $685 \pm 261$ in the sagittal direction. The number of target positions in the sagittal direction was twice as many as in the coronal direction.
It was suggested that the number of pancreatic particles in the driven slice had a significant impact on accuracy.
\par Table~\ref{tbl:multi_metric_vals} shows that the results of ``axial only,'' in which the proposed framework used the correct position from the GT, were $error: 4.26\pm1.16$ and $dice: 0.610\pm0.0519$. By contrast, Table~\ref{tbl:metric_vals} shows that the results of ``psdl,'' in which the proposed framework used the affine parameters that are used when generating the GT, were $error: 5.11\pm1.09$ and $dice: 0.591\pm0.0631$.
The ``axial only'' result was more accurate than the result of ``psdl.'' This suggests that the accuracy of the slice-to-slice registration has a significant impact on the framework.
In addition, the results obtained using multiple slice directions were more accurate than the elastix results ($error: 3.34\pm1.46, dice: 0.646\pm0.114$) described in Section~\ref{sec:vs_elastix}.
In this validation, because we used the correct GT positions for the target position of the selected slices, the conditions were favorable to the proposed framework. Despite these cases, the proposed framework achieved an accuracy comparable to that of volume-to-volume registration, and the result from Section~\ref{sec:vs_elastix} was further reinforced by this.

\section{CONCLUSION}
For the purpose of 3D organ motion estimation from 2D slices regardless of periodicity, we propose a 2D slice-driven 3D organ motion estimation framework and a regression model for selecting the optimal condition of the framework.
First, we confirmed a significant improvement in accuracy by considering the surrounding organs in addition to the target organ, and considering the issue of contact between them.
Second, we compared the accuracy of the proposed framework to that of volume-to-volume registration.
Despite a gap between the proposed framework and volume-to-volume registration, we confirmed that, in some cases, the proposed framework yielded even more accurate results than volume-to-volume registration.
Third, we created regression models and confirmed that the R2 scores were greater than 0.9, and that the RMSEs were sufficiently small for both metrics.
Using these regression models, the proposed framework is able to select the optimal conditions and easily maximize accuracy.
Finally, we confirmed a significant improvement in accuracy using slices in multiple directions. The proposed framework used with multiple slices was comparable to volume-to-volume registration in terms of accuracy.
\par  Since this was an initial study to confirm the feasibility of the proposed method, there are still some key limitations to the proposed framework. The first is ``reliability for real data.'' Because this study aimed to confirm the feasibility of the proposed framework and test our hypothesis regarding the consideration of surrounding organs, we used an open-source dataset and generated GT by geometric deformation and simulation. For practical use, it is necessary to verify the data with images from the same patient taken at different times, as well as with images taken using MR-Linac.  The second is ``simulation time.'' As mentioned in Section~\ref{sec:various_organs}, the proposed framework cannot conduct real-time estimation. We need to further investigate the calculation efficiency based on the GPU used and the trade-off between the reduction of computational complexity by coarsening the discretization and the accuracy of the simulation. The third   is the ``accuracy of the slice-to-slice registration (process 4. in Fig.~\ref{fig:concept}).''
 The results of Section~\ref{sec:multi} suggested that the accuracy of slice-to-slice registration was important. The proposed framework may be vulnerable to motion in perpendicular directions to the slice, if only one slice is used. Because slice-to-slice registration is only needed to compute the target position of each particle in this framework (process 5. in Fig~\ref{fig:concept}), it can use any other slice-to-slice registration methods, or any of the methods mentioned in Section~\ref{sec:intro}. The use of these methods solves the problems of accuracy and motion perpendicular to the slice in process 5., and may further improve overall accuracy.
\par In this study, we focused on MR-Linac radiotherapy; however, the proposed framework could also be used for other types of treatment such as image fusion, motion correction, and volume reconstruction. In the future, we plan to apply our proposed improvement measures and to verify the effectiveness of the method based on data obtained from the device during actual treatment.

\section*{ACKNOWLEDGMENT}
The authors declare the following financial interests/personal
Relationships, which may be considered potential competing interests
with the work reported in this paper:  KJ reports research funding from Elekta. K. K. outside the submitted work. 
\par The authors would like to thank Editage (www.editage.com) for English language editing.

\vfill


\begin{thebibliography}{1}
\bibliographystyle{IEEEtran}

\bibitem{cancer_statistics}R. L. Siegel, K. D. Miller, H. E. Fuchs and A. Jemal, ``Cancer statistics, 2021,'' \textit{CA Cancer J. Clin.}, vol. 71, no. 1, pp. 7-33, Jan. 2021.
\bibitem{rudra} S. Rudra, N. Jiang, S. A. Rosenberg, J. R. Olsen, M. C. Roach, L. Wan, L. Portelance, et al, ``Using adaptive magnetic resonance image-guided radiation therapy for treatment of inoperable pancreatic cancer,'' \textit{Cancer Med.}, vol. 8, no. 5, pp. 2123-2132, May. 2019.
\bibitem{mr_linac}D. Winkel, G. H. Bol, P. S. Kroon, B. van Asselen, S. S. Hackett, A. M. Werensteijn-Honingh, et al, ``Adaptive radiotherapy: The elekta unity MR-linac concept,'' \textit{Clin. Transl. Radiat. Oncol.}, vol. 18, pp. 54\-59, Sep. 2019.
\bibitem{ferrante}E. Ferrante and N. Paragios, ``Slice-to-volume medical image registration: A survey,'' \textit{Med. Image Anal.}, vol. 39, pp. 101-123, Jul. 2017.
\bibitem{paganelli1}C. Paganelli, D. Lee, J. Kipritidis, B. Whelan, P. B. Greer, G. Baroni, et al, ``Feasibility study on 3D image reconstruction from 2D orthogonal cine-MRI for MRI-guided radiotherapy,'' \textit{J. Med. Imaging Radiat. Oncol.}, vol. 62, no. 3, pp. 389-400, Jun. 2018.
\bibitem{seregni}M. Seregni, C. Paganelli, J. Kipritidis, G. Baroni, P. Keall and M. Riboldi, ``Out-of-plane motion correction in orthogonal cine-MRI registration,''  \textit{Radiother. Oncol.}, vol. 123, pp. S147\-S148, May. 2017.
\bibitem{stemkens1} B. Stemkens, R. H. N. Tijssen, B. D. de Senneville, J. J. W. Lagendijk and C. A. T. van den Berg, ``Image-driven, model-based 3D abdominal motion estimation for {MR}-guided radiotherapy,'' \textit{Phys. Med. Biol.}, vol. 61, no. 14, pp. 5335-5355, Jul. 2016.
\bibitem{harris1} W. Harris, L. Ren, J. Cai, Y. Zhang, Z. Chang, and F.-F. Yin, ``A technique for generating volumetric cine-magnetic resonance imaging,'' \textit{Int. J. Radiat. Oncol. Biol. Phys.}, vol. 95, no. 2, pp. 844\-853, Jun. 2016.
\bibitem{harris2} W. Harris, F.-F. Yin, J. Cai, and L. Ren, ``Volumetric cine magnetic resonance imaging (vc-MRI) using motion modeling, free-form deformation and multi-slice undersampled 2d cine MRI reconstructed with spatio-temporal low-rank decomposition,'' \textit{Quant. Imaging. Med. Surg.}, vol. 10, pp. 432-450, Feb. 2020.
\bibitem{garau} N. Garau, R. Via, G. Meschini, D. Lee, P. Keall, M. Riboldi, et al, ``A ROI-based global motion model established on 4DCT and 2D cine-MRI data for MRI-guidance in radiation therapy,'' \textit{Phys. Med. Biol.}, vol. 64, no. 4, pp. 045002, Feb. 2019.
\bibitem{paganelli2}C. Paganelli, S. Portoso, N. Garau, G. Meschini, R. Via, G. Buizza, et al, ``Time-resolved volumetric MRI in MRI-guided radiotherapy: an in silico comparative analysis,'' \textit{Phys. Med. Biol.}, vol. 64, no. 18, pp. 185013, Sep. 2019.
\bibitem{romaguera1}L. V. Romaguera, T. Mezheritsky, R. Mansour, W. Tanguay, and S. Kadoury, ``Predictive online 3d target tracking with population-based generative networks for image-guided radiotherapy,'' \textit{Int. J. Comput. Assist Radiol. Surg.}, vol. 16, no. 7, pp. 1213-1225, Jul. 2021.
\newpage
\bibitem{romaguera2}L. V. Romaguera, T. Mezheritsky, R. Mansour, J.-F. Carrier and S. Kadoury, ``Probabilistic 4D predictive model from in-room surrogates using conditional generative networks for image-guided radiotherapy,'' \textit{Med. Image Anal.}, vol.74, pp. 102250, Dec. 2021.
\bibitem{lombardo}E. Lombardo, M. Rabe, Y. Xiong, L. Nierer, D. Cusumano, L. Placidi, et al, ``Offline and online LSTM networks for respiratory motion prediction in MR-guided radiotherapy,'' \textit{Phys. Med.}, vol. 67, no. 9, pp. 095006, Apr. 2022.
\bibitem{pham1}J. Pham, W. Harris, W. Sun, Z. Yang, F.-F. Yin and L. Ren, ``Predicting real-time 3D deformation field maps (DFM) based on volumetric cine MRI (VC-MRI) and artificial neural networks for on-board 4D target tracking: a feasibility study,'' \textit{Phys. Med. Biol.}, vol. 64, no. 16, pp. 165016, Aug. 2019.
\bibitem{trivisonne}R. Trivisonne, I. Peterlik, S. Cotin and H. Courtecuisse, ``3D physics-based registration of 2D dynamic MRI data,'' \textit{Stud. Health Technol. Inform.}, vol. 220, pp. 432-438, Apr. 2016.
\bibitem{courtecuisse}H. Courtecuisse, Z. Jiang, O. Mayeur, J. F. Witz, P. Lecomte-Grosbras, M. Cosson, et al, ``Three-dimensional physics-based registration of pelvic system using 2D dynamic magnetic resonance imaging slices,'' \textit{Strain}, vol. 56, no. 3, pp. e12339, Apr. 2020.
\bibitem{mpm_book}X. Zhang, Z. Chen, and Y. Liu, ``The material point method: A continuum-based particle method for extreme loading cases,'' \textit{Academic Press}, 2017.
\bibitem{mpm_snow}A. Stomakhin, C. Schroeder, L. Chai, J. Teran, and A. Selle, ``A material point method for snow simulation,'' \textit{ACM Trans. Graph.}, 
vol. 32, no. 102, pp. 1-10, Jul. 2013.
\bibitem{amos}Y. Ji, H. Bai, J. Yang, C. Ge, Y. Zhu, R. Zhang, et al, ``Amos: A large-scale abdominal multi-organ benchmark for versatile medical image segmentation,'' 2022. [Online]. Available: arXiv:2206.08023.
\bibitem{Stomakhin} A.Stomakhin, R. Howes, C. Schroeder and J.Teran, ``Energetically consistent invertible elasticity,'' \textit{Eurographics/ACM SIGGRAPH Symp. on Comp. Anim.}, pp. 25\-32, Jul. 2012.
\bibitem{pancreas} M. Sugimoto, S. Takahashi, M. Kojima, N. Gotohda, Y. Kato, S. Kawano, et al, ``What is the nature of pancreatic consistency? Assessment of the elastic modulus of the pancreas and comparison with tactile sensation, histology, and occurrence of postoperative pancreatic fistula after pancreaticoduodenectomy.'' \textit{Surgery}, vol. 156, no. 5, pp. 1204--1211, Nov. 2014.
\bibitem{stomach_kidney}K. K. Brock, M. B. Sharpe, L. A. Dawson, S. M. Kim and D. A. Jaffray, ``Accuracy of finite element model-based multi-organ deformable image registration,'' \textit{Med. Phys.}, vol. 32, no. 6, pp. 1647-1659, Jun. 2005.
\bibitem{duodenum}B. Hari, S. Bakalis and P. Frye, ``Computational modelling and simulation of the human duodenum,'' \textit{Proc. COMSOL}, 6 pages, Oct. 2012.
\bibitem{elastix}S. Klein, M. Staring, K. Murphy, M. A. Viergever and J. P. Pluim, ``Elastix: A toolbox for intensity-based medical image registration,'' \textit{IEEE Trans. Med. Imaging}, vol. 29, no.1, pp. 196-205, Jan. 2010.
\bibitem{mpm_gpu}M. Gao, X. Wang, K. Wu, A. Pradhana, E. Sifakis, C. Yuksel, et al, ``GPU optimization of material point methods,'' \textit{ACM Trans. Graph.}, vol. 37, no. 6, pp. 1\-12, Dec. 2018.
\bibitem{xgb}T. Chen, C. Guestrin, ``XGBoost: A Scalable Tree Boosting System,'' \textit{Proc. SIGKDD}, pp. 785\-794, Aug. 2016.
\bibitem{optuna} T. Akiba, S. Sano, T. Yanase, T. Ohta and M. Koyama, ``Optuna: A next-generation hyperparameter optimization framework,'' \textit{Proc. SIGKDD}, pp.2623-2631, Jul. 2019.
\end{thebibliography}
\end{document}